# Testing, Evaluation, Verification and Validation (TEVV) of Digital Twins: A Comprehensive Framework

Gabriella Waters
Center for Equitable AI & Machine Learning Systems
Morgan State University

## Abstract
Digital twins have emerged as a powerful technology for modeling and simulating complex systems across various domains (Fuller et al., 2020; Tao et al., 2019). As virtual representations of physical assets, processes, or systems, digital twins enable real-time monitoring, predictive analysis, and optimization. However, as digital twins become more sophisticated and integral to decision-making processes, ensuring their accuracy, reliability, and ethical implementation is essential. This paper presents a comprehensive framework for the Testing, Evaluation, Verification and Validation (TEVV) of digital twins to address the unique challenges posed by these dynamic and complex virtual models.

## Introduction
The concept of digital twins has evolved from its origins in NASA's Apollo missions to become a cornerstone technology of the Fourth Industrial Revolution (Jones et al., 2020). These sophisticated virtual representations leverage real-time data streams, advanced modeling techniques, and artificial intelligence to create dynamic mirrors of physical systems that enable unprecedented insights into system behavior, performance optimization, and predictive maintenance capabilities across industries ranging from manufacturing and healthcare (Sun et al., 2022) to urban planning and aerospace (Barricelli et al., 2019; Liu et al., 2021).

Despite the promising applications and rapid market growth, the deployment of digital twins faces significant challenges related to trust, reliability, and validation. Current digital twin implementations often lack systematic approaches to effectively evaluate their performance (Thelen et al., 2022). There can be uncertainty when attempting to validate virtual models' ability to adequately represent the complexity and dynamic behavior of their physical counterparts. This challenge is particularly acute given that digital twins are increasingly used for critical decision-making processes where errors can have substantial economic, safety, or operational consequences.

The absence of standardized TEVV methodologies creates several critical issues. Many organizations lack guidance on how to systematically assess the quality and reliability of their digital twin implementations. The interoperability between digital twins from different vendors or domains is limited due to inconsistent validation approaches. The ethical implications of digital twin deployment, including harmful algorithmic bias, privacy concerns, and transparency requirements, are often inadequately addressed. Much of the challenge lies in the dynamic nature of digital twins, which continuously evolve through real-time data integration, presents unique validation challenges that traditional modeling approaches cannot always address.

## Current TEVV Challenges in Digital Twin Development

The verification and validation of digital twins presents unique challenges that distinguish them from traditional modeling and simulation approaches (Hua et al., 2022; Somers et al., 2022). Unlike static models, digital twins must maintain fidelity to their physical counterparts while adapting to changing conditions, incorporating new data streams, and supporting real-time decision-making. This dynamic nature creates several key challenges for TEVV practitioners.

Data Quality and Integrity Challenges: Digital twins rely on continuous streams of data from sensors, IoT devices, and external systems, making data quality assessment a vital but complex undertaking (Deantoni et al., 2025). Traditional validation approaches often assume static datasets, but digital twins must handle incomplete, noisy, or contradictory data while maintaining model accuracy. The challenge is compounded by the need to validate data provenance, detect anomalies, and ensure data security throughout the digital twin lifecycle. An example of the digital twin lifecycle can be seen in figure 1.

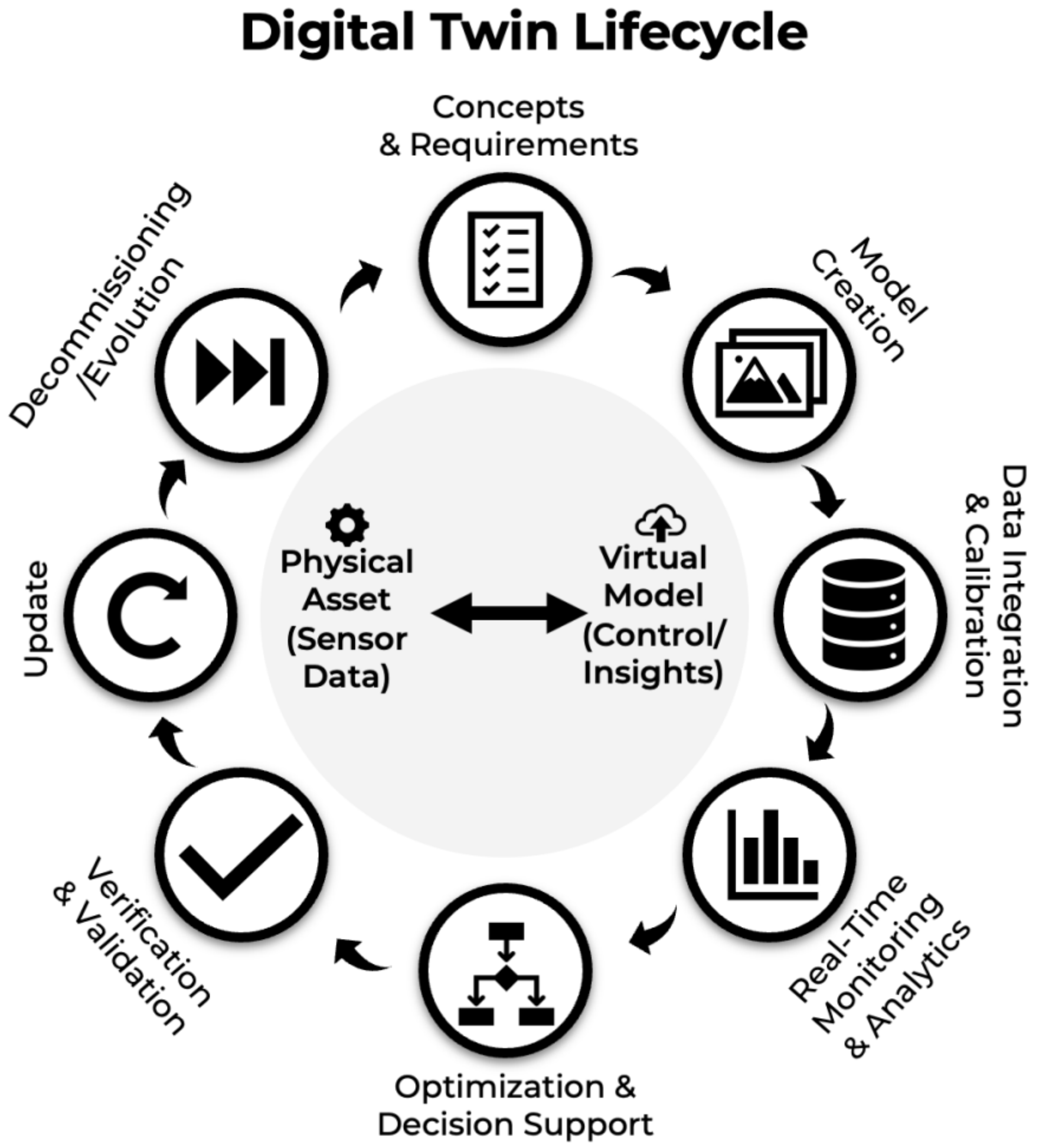

*Figure 1: Digital Twin Lifecycle Loop highlighting the continuous feedback between a physical asset and its virtual model across seven phases, from concept to decommissioning.*

Model Complexity and Uncertainty: Digital twins often integrate multiple modeling approaches, including physics-based models, data-driven machine learning algorithms, and hybrid techniques (Kapteyn et al., 2020; McClellan et al., 2022). This complexity makes it difficult to isolate sources of uncertainty and validate individual model components while ensuring the overall system behavior remains accurate. The interaction between different modeling paradigms introduces emergent behaviors that are challenging to predict and validate using traditional approaches.

Real-time Validation Requirements: The operational nature of digital twins demands continuous validation as they process new data and update their internal states Ding & Xing, 2025). Traditional validation approaches that rely on offline testing and batch processing are insufficient for systems that must maintain accuracy while operating in real-time. This requirement necessitates new approaches to online validation that can detect model degradation, data drift, and performance issues without interrupting operational use.

Interoperability and Standardization Gaps: The lack of standardized approaches to digital twin development has created a fragmented landscape where validation approaches vary significantly across vendors, domains, and applications. This fragmentation limits the ability to compare digital twins, share validation methodologies, and ensure consistent quality standards across different implementations.

**Objectives and Contributions**

This paper addresses the critical need for systematic TEVV approaches in digital twin development by proposing a comprehensive framework that balances methodological rigor with practical implementation considerations. The primary objectives include:

- Development of a Unified TEVV Framework: A structured approach that encompasses testing, evaluation, verification, and validation processes specifically designed for the unique characteristics of digital twins.
- Standardization of Ontological Foundations: Domain-specific ontologies that facilitate consistent representation and validation of digital twins across different application areas.
- Integration of Ethical Considerations: Incorporation of privacy, fairness, transparency, and accountability requirements into the TEVV process to ensure responsible digital twin deployment.
- Practical Implementation Guidance: Actionable methodologies, metrics, and tools that organizations can apply to validate their digital twin implementations.
- Empirical Validation through Case Studies: Demonstration of the framework's applicability through comprehensive examples across multiple domains, including manufacturing, healthcare, financial services, and urban planning.

The key contributions of this work include: (1) a comprehensive TEVV framework specifically designed for digital twins that addresses their dynamic, multi-modal nature; (2) standardized ontological representations that enable consistent validation approaches across domains; (3) integration of ethical considerations as core components of the validation process; (4) practical implementation guidelines that organizations can adapt to their specific contexts; and (5) empirical validation through diverse case studies that demonstrate the framework's effectiveness and versatility.

**Paper Structure and Methodology**

This paper presents the TEVV framework through a systematic progression from theoretical foundations to practical implementation. Following this introduction, the paper examines the fundamental characteristics of digital twins and establishes a taxonomy that informs the subsequent TEVV approaches. Domain-specific ontologies are then presented for manufacturing, healthcare, financial, urban planning, and aerospace applications to provide standardized vocabularies for validation processes.

The core TEVV methodology is presented through four interconnected dimensions: testing approaches that encompass unit, integration, system, and simulation testing; evaluation methodologies covering performance, usability, and value assessment; verification processes for requirements, data, models, and behavior; and validation techniques including empirical, predictive, operational, and conceptual validation. Standardization approaches are then discussed, which provide implementation guidelines and certification processes.

Ethical considerations are integrated throughout the framework to address privacy protection, algorithmic fairness, transparency, and accountability requirements. The practical applicability of the framework is demonstrated through detailed case studies across multiple domains, followed by a discussion of limitations, broader impacts, and future research directions.

**Types of Digital Twins**

Digital twins can be categorized based on their complexity, scope, and application (Pronost et al., 2021). Understanding these types is important for developing appropriate TEVV strategies. Table1 describes the types of digital twins and their key characteristics.

**Table 1: Classification of Digital Twin Types Based on Complexity, Scope, and Application Domain**

| Type of Twin | Description | Key Characteristics | Example |
|---|---|---|---|
| Component Twins | Most basic form, representing individual parts or components of a larger system. Focus is on simulating behavior and performance of specific elements. | • Narrow scope, focusing on a single component<br>• Often used for monitoring performance and predicting maintenance needs<br>• Relatively simple compared to more comprehensive twins | A digital twin of an aircraft engine turbine blade, monitoring its wear and thermal stress over time. |
| Asset Twins | Represent entire machines or standalone assets. Integrate multiple component twins to create a comprehensive model of a complete asset's behavior and performance. | • Broader scope than component twins, encompassing entire machines or systems<br>• Enable holistic performance monitoring and optimization<br>• Often used for predictive maintenance and operational efficiency improvements | A digital twin of an entire wind turbine, including its blades, generator, and control systems, used to optimize power output and maintenance schedules. |
| System or Unit Twins | Model interconnected assets or units working together as part of a larger process or production line. Capture the interactions and dependencies between different assets. | • Model complex interactions between multiple assets<br>• Used for optimizing system-level performance and efficiency<br>• Enable scenario testing and process improvements | A digital twin of a manufacturing production line, simulating the interplay between various machines, material flows, and quality control processes. |
| Process Twins | Focus on modeling entire operational processes, including both physical assets and human interactions. Capture the flow of materials, information, and activities across an organization. | • Encompass both technological and human elements of processes<br>• Used for process optimization, training, and scenario planning<br>• Often integrate with business systems like ERP and MES | A digital twin of a hospital's emergency department, modeling patient flow, resource allocation, and staff interactions to improve efficiency and patient care. |
| Enterprise Twins | Most comprehensive, representing entire organizations or complex systems of systems. Integrate multiple process twins and can model an organization's operations, supply chains, and market interactions. | • Highest level of complexity and scope<br>• Used for strategic decision-making and long-term planning<br>• Often incorporate AI and advanced analytics for predictive insights | A digital twin of an entire smart city, integrating models of transportation systems, energy grids, public services, and citizen behavior to optimize urban planning and resource management. |

## Taxonomy of Digital Twin Characteristics

To effectively assess digital twins, it is essential to establish a taxonomy of their key characteristics. The taxonomy in Table 2 provides a framework for evaluating digital twins across different domains and applications.

**Table 2. Hierarchical Taxonomy Framework for Evaluating Digital Twin Characteristics Across Domains**

| Characteristic | Level | Description |
|---|---|---|
| Model Fidelity and Accuracy | Low | Basic representation with limited accuracy |
| | Medium | Moderate level of detail and accuracy |
| | High | Highly detailed and accurate representation |
| Temporal Resolution | Static | Represents a fixed point in time |
| | Periodic | Updates at regular intervals |
| | Real-time | Continuously updates based on live data |
| Scope of Representation | Micro | Focuses on individual components or processes |
| | Meso | Represents subsystems or limited interactions |
| | Macro | Models entire systems or complex interactions |
| Interactivity | Passive | Provides information without user interaction |
| | Reactive | Responds to user inputs or queries |
| | Proactive | Suggests actions or makes autonomous decisions |
| Data Integration and Synchronization | Unidirectional | Receives data from physical counterpart |
| | Bidirectional | Exchanges data with physical counterpart |
| | Multidirectional | Integrates data from multiple sources |
| Predictive Capability | Descriptive | Represents current state |
| | Diagnostic | Analyzes causes of events or behaviors |
| | Predictive | Forecasts future states or outcomes |
| | Prescriptive | Recommends actions based on predictions |
| Adaptability | Fixed | Maintains a static model |
| | Configurable | Allows manual adjustments to parameters |
| | Self-learning | Automatically updates based on new data |
| Interoperability | Isolated | Functions as a standalone system |
| | Partially Integrated | Interfaces with some other systems |
| | Fully Integrated | Seamlessly integrates with multiple systems and digital twins |
| Usability and Visualization | Basic | Limited user interface and visualization capabilities |
| | Intermediate | User-friendly interface with standard visualization tools |
| | Advanced | Intuitive interface with advanced 3D visualization and interaction capabilities |
| Adaptability and Learning Capability | Fixed | Maintains a static model |
| | Configurable | Allows manual adjustments to parameters |
| | Self-Learning | Automatically updates and improves based on new data |
| Security and Privacy | Low | Minimal data protection and access controls |
| | Medium | Industry-standard security protocols implemented |
| | High | Robust encryption, access controls, and privacy safeguards |
| Validation Against Physical Twin | Low | Limited comparison to physical twin data |
| | Medium | Periodic validation against key metrics |
| | High | Continuous validation across multiple parameters |

*Table 2 presents a hierarchical taxonomy used to evaluate the characteristics of digital twins in multiple application domains, and organizes key features and assessment criteria to enable systematic comparison and analysis.*

## Ontologies for Digital Twins

Ontologies play a key role in establishing a common language and structure for digital twins across different domains (Karabulut et al., 2023; Wilson et al., 2024). They define the relationships between various elements of a digital twin and provide a framework for interoperability and knowledge sharing.

### Core Digital Twin Ontology

The Core Digital Twin Ontology (CDTO) provides a foundation for describing the fundamental components and relationships common to all digital twins (Wilson et al., 2024):

- Physical Entity: The real-world object or system being modeled
- Virtual Entity: The digital representation of the physical entity
- Data: Information collected from sensors, historical records, or simulations
- Model: Mathematical or logical representation of the system's behavior
- Interface: Methods for interaction between the physical and virtual entities
- Simulation: Process of running the model to predict outcomes
- Visualization: Representation of data and simulation results

### Domain-Specific Ontologies

Building on the CDTO, domain-specific ontologies can be developed to address the unique requirements of different fields. Several domains are listed below as examples.

### Manufacturing Digital Twin Ontology

The Manufacturing Digital Twin Ontology extends the Core Digital Twin Ontology to capture the complexities of modern manufacturing environments (Qi et al.,2021; Cimino et al., 2019). Recent systematic reviews highlight the unique verification and validation requirements for manufacturing digital twins, and emphasizes the need for domain-specific metrics, continuous validation, and integration of data-driven and physics-based approaches (Bitencourt et al., 2024; Kapteyn et al., 2020). Physics-based modeling remains foundational for digital twins to enable accurate simulation of complex systems and supporting real-time control and optimization (Ammar et al., 2024; Ding & Xing, 2025). It encompasses key aspects such as:

Production Process: Models the step-by-step procedures involved in creating products. For example, it might include concepts like "Assembly Line", "Batch Processing," and "Just-In-Time Manufacturing." A specific instance could be "Automotive_Assembly_Line_A1" with properties like cycle time, throughput, and current status.

Equipment: This section represents the machinery and tools used in manufacturing. It includes classes for different types of equipment (e.g., "CNC Machine," "Robotic Arm," "Conveyor Belt") along with their specifications, operational parameters, and maintenance schedules. An example instance might be "Welding_Robot_W23" with properties for welding speed, accuracy, and remaining electrode life.

Material Flow: Models the movement of raw materials, work-in-progress, and finished goods through the factory. It includes concepts like "Inventory," "Workstation," and "Material_Handling_System_A." An example could be tracking the flow of steel sheets from the raw material storage through various processing stations to become car body panels.

Quality Control: This section represents the processes and metrics used to ensure product quality. It includes concepts like "Inspection Point," "Quality Metric," and "Defect Classification." An instance might be "Final_Inspection_Station_F3" with associated pass/fail criteria and historical defect rates.

Supply Chain: Models the external connections of the manufacturing process, including suppliers, logistics, and distribution networks. It could include concepts like "Supplier," "Transportation Mode," and "Distribution Center." An example instance might be "Supplier_XYZ" with properties for lead times, reliability scores, and current order status.

Healthcare Digital Twin Ontology

The Healthcare Digital Twin Ontology adapts the core concepts to the medical field, focusing on patient care and healthcare operations (Katsoulakis et al., 2024; Venkatesh et al., 2022). Digital twins are increasingly being deployed in precision health applications, where they enable individualized treatment planning, disease prediction, and population-level health interventions (Bian et al., 2024; Shen et al., 2024):

Patient: Represents individual patients and their health status. It includes concepts like "Medical History," "Vital Signs," and "Genetic Profile." An example instance could be "Patient_12345" with properties for current medications, allergies, and recent lab results.

Physiological Systems: Models the various systems of the human body. It includes concepts like "Cardiovascular System," "Respiratory System," and "Nervous System." An example could be a digital twin of a patient's heart, modeling parameters like heart rate, blood pressure, and ejection fraction.

Treatment Protocols: Represents standardized medical procedures and treatment plans. It includes concepts like "Medication Regimen," "Surgical Procedure," and "Rehabilitation Program." An instance might be "Chemotherapy_Protocol_A" with associated drug dosages, treatment schedule, and expected outcomes.

Medical Devices: Models the equipment used in patient care and monitoring. It includes concepts like "Ventilator," "MRI Machine," and "Infusion Pump." An example could be "Ventilator_V101" with real-time data on settings, alarms, and patient response.

Healthcare Facility: Represents the physical and operational aspects of healthcare institutions. It includes concepts like "Hospital Ward," "Operating Room," and "Emergency Department." An instance might be "General_Ward_3B" with data on bed occupancy, staffing levels, and current patient acuity.

Financial/Transactional Digital Twin Ontology
The Financial/Transactional Digital Twin Ontology extends the Core Digital Twin Ontology to model complex financial systems and transactions. It encompasses key aspects such as:

Financial Entity: Represents the actors in the financial system. It includes concepts like "Individual, "Corporation," and "Financial Institution." An example instance could be "Bank_XYZ" with properties like total assets, risk profile, and regulatory compliance status.

Transaction: Models individual financial exchanges. It includes classes for different types of transactions (e.g., "Payment," "Investment," "Loan") along with their attributes and relationships. An example instance might be "Transaction_ID_12345" with properties for amount, timestamp, sender, receiver, and associated financial products.

Financial Product: Represents various financial instruments and services. It includes concepts like "Savings Account," "Mortgage," and "Stock Option." An instance could be "MortgageProduct_A" with properties for interest rate, term length, and associated risks.

Market: Models the environment where financial transactions occur. It includes concepts like "Stock Market," "Foreign Exchange Market," and "Commodities Market." An example could be "NYSE_TwinInstance" with real-time data on trading volume, market indices, and current volatility.

Regulatory Framework: Represents the rules and regulations governing financial activities. It includes concepts like "Banking Regulation," "Securities Law," and "Anti-Money Laundering Policy." An instance might be "Dodd-Frank_Act" with associated compliance requirements and affected financial entities.

Economic Indicator: Models key metrics that reflect economic health and trends. It includes concepts like "GDP," "Inflation Rate," and "Unemployment Rate." An example instance could be "US_InflationRate_2024" with historical data, current value, and forecasted trends.

Risk Model: Represents methodologies for assessing and quantifying financial risks. It includes concepts like "Credit Risk Model," "Market Risk Model," and "Operational Risk Model." An instance might be "CreditScoringModel_V3" with properties for input parameters, calculation methodology, and historical performance.

Fraud Detection System: Represents mechanisms for identifying and preventing fraudulent activities. It includes concepts like "Transaction Monitoring," "Identity Verification," and "Anomaly Detection." An example could be "FraudDetectionSystem_F2" with real-time data on monitored transactions, alert thresholds, and detected anomalies.

Urban Planning Digital Twin Ontology

The Urban Planning Digital Twin Ontology applies digital twin concepts to city management and development (Petrova-Antonova & Ilieva, 2020):

Infrastructure: Models the physical structures and systems that support urban life. It includes concepts like "Building," "Utility Network," and "Public Space." An example instance could be "Water_Treatment_Plant_A" with properties for capacity, current load, and maintenance schedule.

Transportation Networks: Represents the systems for moving people and goods within the urban area. It includes concepts like "Road Network," "Public Transit System," and "Bike Lane." An instance might be "Subway_Line_2" with real-time data on train locations, passenger counts, and service disruptions.

Population Dynamics: Models the characteristics and movements of the city's inhabitants. It includes concepts like "Demographic Group," "Migration Pattern," and "Economic Activity." An example could be modeling the daily commute patterns of different demographic groups.

Environmental Factors: Represents the natural and man-made environmental conditions affecting the city (Hazeleger et al., 2024; Blair, 2025). It includes concepts like "Air Quality," "Green Space," and "Noise Level." An instance might be "City_Park_X" with data on vegetation health, visitor numbers, and local microclimate.

Zoning Regulations: Models the legal and administrative framework governing urban development. It includes concepts like "Land Use Classification," "Building Code," and "Historic Preservation Area." An example could be a digital representation of zoning changes and their projected impact on traffic patterns and property values.

Aerospace Digital Twin Ontology

The Aerospace Digital Twin Ontology tailors the digital twin concept to aircraft design, operation, and maintenance:

Aircraft Components: Represents the physical parts and systems of an aircraft. It includes concepts like "Airframe," "Engine," and "Avionics System." An example instance could be "Engine_E1" of a specific aircraft, with real-time data on performance parameters, wear indicators, and maintenance history.

Flight Dynamics: Models the behavior of the aircraft in flight. It includes concepts like "Aerodynamics," "Propulsion," and "Flight Control." An instance might be a simulation of the aircraft's performance under various atmospheric conditions and flight maneuvers.

Maintenance Procedures: Represents the processes for keeping the aircraft in optimal condition. It includes concepts like "Inspection Schedule," "Repair Procedure," and "Part Replacement." An example could be "Annual_Airframe_Inspection" with associated checklists, required equipment, and historical findings.

Environmental Conditions: Models the external factors affecting aircraft operation. It includes concepts like "Weather Pattern," "Atmospheric Turbulence," and "Runway Condition," An instance might be a real-time model of weather conditions along a flight route, including wind speeds, precipitation, and visibility.

Air Traffic Control: Represents the systems and procedures for managing air traffic. It includes concepts like "Flight Plan," "Airspace Sector," and "Communication Protocol." An example could be a digital twin of an air traffic control center that models current traffic patterns, sector workloads, and potential conflicts.

## TEVV Approaches for Digital Twins

The TEVV process for digital twins must address the unique challenges posed by their dynamic nature and complex interactions (Lugaresi et al., 2023). This section outlines key approaches for ensuring the accuracy, reliability, and effectiveness of digital twins.

### Testing

Testing is an essential phase in the development and maintenance of digital twins to make sure that they function as intended and meet specified requirements (Somers et al., 2022). It involves a systematic process of examining various aspects of the digital twin to detect errors, assess performance, and verify behavior under different conditions.

Testing digital twins can be challenging due to their complex nature that often involves real-time data processing, advanced analytics, and intricate interactions with physical systems. The testing process typically encompasses multiple levels, from examining individual components to assessing the entire system's functionality. It also includes testing in simulated environments to explore scenarios that might be impractical or dangerous to test in real-world settings.

Effective testing of digital twins requires a combination of traditional software testing methodologies and specialized approaches that account for the unique characteristics of cyber-physical systems (Flammini, 2021). This process is important for building confidence in the digital twin's reliability, accuracy, and robustness, ultimately verifying that it can serve as a trustworthy simulation/virtual representation of its physical counterpart (Lu et al., 2024).

Unit Testing:
Unit testing is a fundamental approach in digital twin development that focuses on verifying individual components or modules (Somers et al., 2022). This process involves rigorous testing of small, isolated parts of the digital twin to confirm that it functions correctly. For example, a unit test might check if a sensor data processing module accurately converts raw input into meaningful metrics. Methods include input validation to ensure the component handles various data types correctly, output verification to confirm expected results, boundary value analysis to test edge cases, and error handling checks to ensure robustness under unexpected conditions.

Integration Testing:
Integration testing examines how different components of the digital twin work together. This stage is necessary for identifying issues that may arise when individual modules interact (Wagg et al., 2020). For instance, it might test how a data collection component interfaces with a machine learning prediction module. Methods include interface testing to verify correct data exchange between components, data flow analysis to track information through the system, dependency validation to ensure components interact in the right order, and synchronization checks to confirm real-time components operate in harmony.

System Testing:
System testing evaluates the digital twin as a complete entity, including its interactions with external systems (Lu et al., 2024). This comprehensive approach certifies that the digital twin functions as intended in a real-world context. This could involve testing how the digital twin of a manufacturing plant interacts with enterprise resource planning software. Methods include end-to-end scenario testing to simulate real-world use cases, performance testing under various loads to ensure scalability, compatibility testing with different platforms to ensure broad usability, and security testing to protect against potential vulnerabilities.

Simulation Testing:

Simulation testing assesses the digital twin's behavior in simulated environments to allow for the exploration of scenarios that might be impractical or dangerous to test in reality (Somers et al., 2022). An example of this could involve simulating extreme weather conditions for a digital twin of an offshore wind farm. Methods include Monte Carlo simulations to test performance across a range of random inputs, scenario-based testing to evaluate specific use cases, sensitivity analysis to understand how changes in inputs affect outputs, and extreme condition testing to assess behavior under unlikely but critical situations.

**Evaluation**

Evaluation of digital twins is a comprehensive assessment process that goes beyond basic functionality testing to determine the overall quality, effectiveness, and value of the digital twin in meeting its intended objectives. This process examines various aspects of the digital twin's performance, usability, and impact on operations or decision-making processes.

Evaluation is essential because it provides insights into how well the digital twin fulfills its purpose in real-world scenarios. It considers factors such as accuracy of predictions, responsiveness, resource efficiency, and scalability. Additionally, evaluation assesses the user experience and examines how easily stakeholders can interact with and derive value from the digital twin.

An important aspect of evaluation in many domains is determining the return on investment and overall value proposition of the digital twin. This involves analyzing cost-effectiveness, improvements in operational efficiency, and the digital twin's contribution to risk reduction or enhanced decision-making. Comparative evaluation also plays a key role in benchmarking the digital twin against industry standards or alternative solutions to determine if it represents a best-in-class solution for its intended application.

Performance Evaluation:

Performance evaluation assesses how well the digital twin meets its operational objectives (Tang et al., 2025; Cardoso et al., 2025). Measurements of key performance indicators such as the accuracy of predictions (e.g., how closely the digital twin's output matches real-world data), response time (how quickly it processes inputs and provides outputs), resource utilization (efficiency in using computational resources), and scalability (ability to handle increasing amounts of data or users) are part of this kind of evaluation.

Usability Evaluation:

Usability evaluation focuses on the user experience of interacting with the digital twin (Zare & Lazarova-Molnar, 2025). This includes assessing the ease of interaction (how intuitively users can navigate and operate the digital twin), clarity of visualizations (how effectively data and insights are presented visually), intuitiveness of the interface (how easily users can understand and use features without extensive training), and accessibility for different user groups (ensuring the digital twin can be used by individuals with varying levels of expertise or physical/cognitive abilities).

Utility Evaluation:
Utility evaluation determines whether the digital twin provides the features and functionality that users actually need to accomplish their goals (Blair, 2025). This assessment focuses on the core value proposition of the digital twin to examine whether it solves real problems and meets genuine user requirements. Key aspects include:

- Functional Completeness: Does the digital twin provide all the essential features required for the intended use cases?
- Problem-Solution Fit: Does the digital twin effectively address the specific challenges it was designed to solve?
- Value Delivery: Does the digital twin enable users to achieve their objectives more effectively than alternative approaches?
- Need Satisfaction: Does the digital twin meet both explicit and implicit user requirements across different stakeholder groups?

Utility and usability work together to determine overall usefulness. While usability focuses on how easy and pleasant features are to use, utility addresses whether the system provides the features users need in the first place. A digital twin may be highly usable but lack utility if it doesn't solve real problems, or it may have high utility but poor usability if essential features are difficult to access or operate.

Integrated Utility-Usability Assessment:
The evaluation process should recognize that users will often tolerate poor usability when a system provides high utility - a phenomenon known as the "utility-over-usability effect." This is especially relevant for digital twins in critical domains where the value of insights and predictions may outweigh interface difficulties. However, the goal should be to achieve both high utility and high usability to maximize the digital twin's effectiveness and adoption.

The assessment should include metrics for both dimensions:

- Utility Metrics: Feature completeness scores, problem-solving effectiveness ratings, value realization measurements, and user needs satisfaction indices
- Usability Metrics: Task completion rates, error frequencies, learning curve measurements, and user satisfaction scores
- Combined Usefulness Metrics: Overall system effectiveness, user retention rates, and comparative performance against alternative solutions

Value Assessment:
Value assessment determines the overall benefit and return on investment of the digital twin. Conducting a cost-benefit analysis (comparing the costs of development and operation against the benefits gained), evaluating the impact on decision-making (how the digital twin improves the quality and speed of decisions), assessing improvement in operational efficiency (quantifying productivity gains or cost savings), and measuring risk reduction (how the digital twin helps in identifying and mitigating potential risks) happens at this level.

Comparative Evaluation:
Comparative evaluation places the digital twin in context by comparing it to alternatives or standards (Cardoso et al., 2025). This incorporates benchmarking against industry standards (how the digital twin performs compared to established benchmarks in the field), comparison with alternative solutions (how it stacks up against other approaches to solving the same problem), and historical performance analysis (how the digital twin's performance has improved over time or compared to previous versions).

**Verification**

Verification in digital twin development is the process of ensuring that the digital twin is built correctly and meets its specified requirements and design criteria (Huang et al., 2024). This process focuses on confirming that the implementation of the digital twin aligns with its intended design and that all components work together as planned.

Verification is a systematic and often formal process that examines various aspects of the digital twin, including its underlying models, data handling processes, and behavioral characteristics. It involves checking that the digital twin's code, algorithms, and mathematical models are correctly implemented and that they produce expected outputs for given inputs.

Key aspects of verification include making sure that the digital twin accurately represents the conceptual model it's based on, that it correctly processes and interprets data, and that it behaves consistently across different operating conditions. Verification may employ techniques such as formal methods, code reviews, and automated testing to rigorously examine the digital twin's implementation.

While validation focuses on whether the right digital twin is built for the intended purpose, verification ensures that the digital twin is built correctly according to its specifications. Together, verification and validation form a comprehensive approach to ensuring the quality and reliability of digital twins.

Requirements Verification:
Requirements verification ensures that the digital twin meets all specified requirements (Shao et al., 2024). Techniques include using a traceability matrix to link requirements to specific features or components, applying formal methods to mathematically prove correctness, employing model checking to verify system properties, and conducting static analysis of the code to identify potential issues before runtime.

Data Verification:
Data verification focuses on ensuring the quality and integrity of data used by the digital twin (Liu et al., 2021). This includes data quality assessment to check for completeness and accuracy, consistency checks to ensure data aligns across different parts of the system, anomaly detection to identify unusual patterns that might indicate errors, and source validation to verify the reliability of data inputs.

Model Verification:
Model verification ensures that the mathematical or logical models underlying the digital twin are correct (Huang et al., 2024). This approach can incorporate mathematical proofs to demonstrate the validity of algorithms, formal verification to check that the model meets its specification, code review to identify potential errors in implementation, and symbolic execution to analyze possible execution paths.

Behavior Verification:
Behavior verification ensures that the digital twin behaves as expected under various conditions (Bitencourt et al., 2024). Elements of this level of verification include state space exploration to examine all possible states of the system, invariant checking to ensure critical properties always hold, temporal logic verification to check behavior over time, and assertion-based verification to test specific conditions during execution.

**Validation**

Validation in the context of digital twins is the process of determining whether the digital twin accurately represents the real-world system it's designed to model and meets the needs of its intended users (Sel et al., 2025). This process is necessary for establishing the credibility and reliability of the digital twin in practical applications.

Validation goes beyond checking if the digital twin works correctly from a technical standpoint; it assesses whether the digital twin provides a meaningful and accurate representation of the physical system under various conditions. This includes verifying that the digital twin's outputs and predictions align with real-world observations and that it can effectively simulate the behavior of the physical system across different scenarios.

The validation process often involves collaboration between technical experts who understand the digital twin's inner workings and domain experts who have deep knowledge of the physical system being modeled. It may include empirical testing, comparison with historical data, expert reviews, and long-term performance monitoring in operational settings. Ultimately, validation ensures that the digital twin can be trusted as a reliable tool for analysis, prediction, and decision-making in its intended application domain.

Empirical Validation:
Empirical validation assesses how well the digital twin represents the real-world system it models (Gao et al., 2025; Hua et al., 2022). Comparison with physical measurements to check accuracy, historical data validation to ensure the model aligns with past observations, expert review to leverage domain knowledge in assessing the model's realism, and field testing to evaluate performance in actual operational conditions are included here.

Predictive Validation:
Predictive validation focuses on the digital twin's ability to forecast future states or behaviors (Zare & Lazarova-Molnar, 2024). This include backtesting to evaluate predictions against historical data, cross-validation to assess performance on unseen data, out-of-sample testing to check generalization to new scenarios and forecast evaluation to measure the accuracy of predictions over time.

Operational Validation:
Operational validation ensures the digital twin meets the needs of its users in real-world settings (Sel et al., 2025) and can employ the following approaches: user acceptance testing to gather feedback from intended users, pilot studies to test the digital twin in limited real-world deployments, operational scenario testing to evaluate performance under typical use cases, and long-term performance monitoring to track reliability and effectiveness over extended periods.

Conceptual Validation:
Conceptual validation assesses whether the digital twin's underlying concepts and assumptions are sound (Bitencourt et al., 2024). This validation includes face validity assessment to gauge initial plausibility, structured walkthroughs to systematically review the model's logic, Delphi techniques to gather consensus from multiple experts, and causal-loop diagram analysis to examine the relationships between different elements of the system.

**Standardization of TEVV Approaches**
While the development and deployment of digital twins may vary across industries and applications, standardizing TEVV approaches can ensure consistency, reliability, and interoperability (Shao, 2024). Table 3 presents a structured overview of the primary categories, subcategories, purposes, and representative methods associated with the Testing, Evaluation, Verification, and Validation (TEVV) of digital twins. This taxonomy is designed to clarify the distinct objectives and techniques at each stage of the TEVV process to enable practitioners to systematically address the functional, operational, and quality assurance needs of digital twin systems.

**Table 3. Standardized Metrics and Key Performance Indicators for Digital Twin TEVV**

| Category | Subcategory | Purpose/Focus | Methods/Techniques |
|---|---|---|---|
| Testing | Unit Testing | Verify individual components | • Input validation<br>• Output verification<br>• Boundary value analysis<br>• Error handling checks |
| | Integration Testing | Ensure proper component interaction | • Interface testing<br>• Data flow analysis<br>• Dependency validation<br>• Synchronization checks |
| | System Testing | Evaluate the digital twin as a whole | • End-to-end scenario testing<br>• Performance testing<br>• Compatibility testing<br>• Security testing<br>• Fuzz testing |
| | Simulation Testing | Assess behavior in simulated environments | • Monte Carlo simulations<br>• Scenario-based testing<br>• Sensitivity analysis<br>• Extreme condition testing |
| Evaluation | Performance Evaluation | Assess operational effectiveness | • Accuracy of predictions<br>• Response time<br>• Resource utilization<br>• Scalability |
| | Usability Evaluation | Assess user experience | • Ease of interaction<br>• Clarity of visualizations<br>• Intuitiveness of interface<br>• Accessibility |
| | Value Assessment | Determine overall benefit and ROI | • Cost-benefit analysis<br>• Impact on decision-making<br>• Operational efficiency improvement<br>• Risk reduction |
| | Comparative Evaluation | Compare to alternatives or standards | • Benchmarking<br>• Comparison with alternatives<br>• Historical performance analysis |
| Verification | Requirements Verification | Ensure meeting of specifications | • Traceability matrix<br>• Formal methods<br>• Model checking<br>• Static analysis |
| | Data Verification | Ensure data quality and integrity | • Data quality assessment<br>• Consistency checks<br>• Anomaly detection<br>• Source validation |
| | Model Verification | Verify underlying models | • Mathematical proof<br>• Formal verification<br>• Code review<br>• Symbolic execution |
| | Behavior Verification | Ensure expected system behavior | • State space exploration<br>• Invariant checking<br>• Temporal logic verification<br>• Assertion-based verification |
| Validation | Empirical Validation | Assess real-world representation | • Comparison with physical measurements<br>• Historical data validation<br>• Expert review, Field testing |
| | Predictive Validation | Evaluate forecasting ability | • Backtesting<br>• Cross-validation<br>• Out-of-sample testing<br>• Forecast evaluation |
| | Operational Validation | Ensure meeting of user needs | • User acceptance testing<br>• Pilot studies<br>• Operational scenario testing<br>• Long-term performance monitoring |
| | Conceptual Validation | Assess underlying concepts | • Face validity assessment<br>• Structured walkthrough<br>• Delphi technique<br>• Causal-loop diagram analysis |

*Table 3 presents a hierarchical taxonomy used to evaluate the characteristics of digital twins in multiple application domains with key features and assessment criteria described to enable systematic comparison and analysis.*

**Standardized TEVV Framework**
A standardized TEVV framework for digital twins should encompass the following key elements (Shao et al., 2024):

1. Verification Protocols

Model Verification:
Establishing protocols for verifying the mathematical and logical correctness of digital twin models is necessary for ensuring their reliability and accuracy. This process involves several key components:

Code review procedures: Implement systematic code reviews where experienced developers examine the model's source code for errors, inefficiencies, and adherence to coding standards. This can help identify logical errors and improve overall code quality.

Unit testing standards: Develop comprehensive unit tests for individual components of the digital twin model. These tests should cover various input scenarios and edge cases to ensure each component functions correctly in isolation.

Integration testing guidelines: Create guidelines for testing how different components of the digital twin interact when assembled. This includes testing data flow between components, checking for unexpected behaviors when components are combined, and verifying that the integrated system meets overall requirements.

Formal methods for critical systems: For digital twins of critical systems (e.g., in aerospace or healthcare) formal verification methods should be employed. These mathematical techniques can prove the correctness of algorithms and ensure the model behaves as specified under all possible inputs.

Data Verification:
Ensuring data quality and integrity is essential for digital twins to provide accurate representations and predictions. Standards for data verification should include:

Data cleaning and preprocessing procedures: Develop standardized methods for identifying and handling missing values, outliers, and inconsistencies in raw data before it's used in the digital twin.

Anomaly detection methods: Implement automated systems to detect unusual patterns or values in incoming data that could indicate sensor malfunctions or other issues.

Data consistency checks: Establish procedures to verify that data remains consistent across different sources and over time, especially for digital twins that integrate data from multiple systems.

Version control and data lineage tracking: Implement robust version control systems for both the digital twin model and its associated data. Maintain clear records of data provenance to understand how data has been transformed and used throughout the digital twin's lifecycle.

2. Validation Methodologies

Empirical Validation:
Standardizing approaches for comparing digital twin outputs with real-world data is critical for ensuring the twin's accuracy and reliability:

Statistical methods for quantifying prediction accuracy: Develop a suite of statistical measures (e.g., mean squared error, R-squared, confusion matrices for classification tasks) to assess how closely the digital twin's predictions match observed data.

Procedures for continuous validation: Establish automated systems that continuously compare digital twin outputs to new real-world data as it becomes available to allow for ongoing assessment of the twin's accuracy.

Guidelines for handling discrepancies: Create clear protocols for investigating and addressing situations where the digital twin's predictions significantly deviate from observed data, including procedures for model recalibration or refinement.

Operational Validation:
Assessing the digital twin's performance in its intended operational context ensures it meets practical needs:

User acceptance testing protocols: Develop standardized procedures for end-users to evaluate the digital twin's functionality, usability, and value in real-world scenarios.

Scenario-based testing procedures: Create a diverse set of operational scenarios that the digital twin must successfully navigate, including both typical use cases and edge cases.

Long-term performance monitoring standards: Establish metrics and procedures for tracking the digital twin's performance over extended periods, ensuring it continues to provide value and accurate insights as the physical system evolves.

3. Evaluation Metrics

Performance Metrics:
Defining standardized metrics for evaluating digital twin performance ensures consistent assessment across different implementations:

Accuracy measures: Adopt widely accepted statistical measures like Mean Absolute Error (MAE) and Root Mean Square Error (RMSE) to quantify prediction accuracy across various aspects of the digital twin.

Computational efficiency metrics: Develop benchmarks for assessing the digital twin's processing speed, memory usage, and scalability to ensure it can operate effectively in real-time or near-real-time environments.

Scalability assessments: Create standardized tests to evaluate how well the digital twin's performance scales with increasing data volumes, complexity of simulations, or number of concurrent users.

Responsiveness to real-time data updates: Establish metrics to measure how quickly and accurately the digital twin incorporates new data and updates its predictions or simulations.

Usability Metrics:
Assessing the digital twin's user interface and experience ensures it can be effectively utilized by its intended users:

Ease of use ratings: Develop standardized usability scales and questionnaires to gather quantitative feedback on the digital twin's interface and overall user experience.

User satisfaction surveys: Create comprehensive surveys to collect detailed qualitative and quantitative feedback on various aspects of the digital twin's functionality and value to users.

Task completion time measurements: Establish standardized tasks and measure the time required for users to complete them, providing objective data on the digital twin's efficiency and ease of use.

Utility Metrics: Evaluate the extent to which the digital twin provides the necessary features, functions, and information needed to achieve user and organizational objectives. Utility metrics answer the question: “Does the digital twin do what users need it to do?” Key measures include:

- Functional completeness: Percentage of required features or use cases supported by the digital twin.
- Problem-solution fit: Degree to which the digital twin addresses the real-world challenges it was designed for, as assessed through stakeholder feedback or use-case coverage.
- Decision support effectiveness: Impact of the digital twin on the quality and speed of decision-making, such as reductions in time-to-decision or improvements in decision outcomes.
- Adoption and engagement rates: Frequency and breadth of use across intended user groups, indicating whether the digital twin is seen as useful and relevant in practice.
- User-reported utility: Direct feedback from users on the perceived usefulness and relevance of the digital twin’s features and outputs.

4. Testing Procedures

Scenario Testing:
Developing standardized scenarios for testing digital twins across different industries ensures comprehensive evaluation:

Normal operating conditions: Create a set of baseline scenarios that represent typical day-to-day operations for the system being modeled, ensuring the digital twin performs accurately under standard conditions.

Edge cases and extreme scenarios: Develop a suite of tests that push the digital twin to its limits, simulating rare but critical situations to ensure it can handle unexpected or extreme conditions.

Failure mode simulations: Design scenarios that simulate various types of system failures or degradations, testing the digital twin's ability to predict, detect, and respond to these situations.

Interoperability Testing:
Creating protocols for ensuring digital twins can interact with other systems is crucial for integration into existing infrastructures:

Data exchange format standards: Establish common data formats and protocols for digital twins to exchange information with other systems, ensuring seamless integration and data flow (Drobnjakovic et al., 2023).

API testing procedures: Develop comprehensive test suites for validating the functionality, security, and performance of APIs used by digital twins to communicate with external systems.

Cross-platform compatibility checks: Create testing procedures to verify that digital twins can operate consistently across different hardware platforms, operating systems, and software environments.

5. Documentation and Reporting

TEVV Documentation:
Standardizing the documentation required for TEVV processes ensures comprehensive and consistent record-keeping:

Test plans and procedures: Develop templates and guidelines for creating detailed test plans that outline the scope, objectives, and methodologies for each phase of TEVV.

Verification and validation reports: Establish standardized formats for reporting the results of verification and validation activities, including clear presentation of metrics, test outcomes, and any identified issues.

Traceability matrices: Create standardized matrices that link each requirement of the digital twin to specific test cases and results to ensure comprehensive coverage and facilitate audits.

Certification Process:
Establishing a standardized certification process for digital twins promotes trust and quality assurance:

Define certification levels: Create a tiered certification system based on the rigor of TEVV processes applied, allowing for different levels of certification depending on the critical nature of the digital twin's application.

Third-party audit framework: Develop guidelines and procedures for independent third-party audits of digital twins to allow for the objective assessment and validation of TEVV processes and outcomes.

Implementation Guidelines
The following approaches are suggested to effectively implement these standardized TEVV approaches:

1. Industry Collaboration:

Engaging with industry stakeholders is crucial to ensure that TEVV standards are practical, relevant, and widely applicable. This collaboration should involve:

Forming industry working groups: Establish cross-sector committees that bring together experts from various fields such as manufacturing, healthcare, aerospace, and IT to contribute to the development of TEVV standards.

Regular workshops and conferences: Organize events where industry professionals can share best practices, discuss challenges, and contribute to the evolution of TEVV standards.

Pilot programs: Implement pilot programs in different industries to test and refine TEVV standards and gather real-world feedback on their effectiveness and practicality.

2. Regulatory Alignment:
Aligning TEVV standards with existing regulatory frameworks is essential for widespread adoption and compliance:

Regulatory mapping: Conduct comprehensive analyses to map how TEVV standards align with existing regulations in various industries (e.g., FDA regulations for medical devices, FAA regulations for aerospace).

Collaborative development: Work closely with regulatory bodies to ensure TEVV standards meet or exceed existing regulatory requirements.

Compliance guidance: Develop clear guidelines to help organizations understand how adhering to TEVV standards supports regulatory compliance in their respective industries.

3. Flexibility:
Designing standards to be flexible enough to accommodate different types of digital twins and evolving technologies is crucial for long-term relevance:

Modular framework: Create a modular TEVV framework that allows organizations to apply relevant components based on their specific digital twin applications.

Technology-agnostic approach: Develop standards that focus on principles and outcomes rather than specific technologies to allow for adaptation as new technologies emerge.

Regular review cycles: Establish a process for periodically reviewing and updating standards to ensure they remain relevant as digital twin technology evolves.

4. Continuous Improvement:
Establishing mechanisms for regularly reviewing and updating TEVV standards based on new research and industry feedback ensures ongoing relevance and effectiveness.

- Feedback loops: Create formal channels for practitioners to provide feedback on the standards' effectiveness and suggest improvements.
- Research partnerships: Collaborate with academic institutions and research organizations to stay abreast of the latest developments in digital twin technology and TEVV methodologies.
- Version control: Implement a clear versioning system for TEVV standards, with scheduled major reviews and updates to incorporate significant advancements or changes in best practices.

5. Training and Education:
Developing standardized training programs for TEVV practitioners ensures consistent application of standards.

- Certification programs: Create professional certification programs for TEVV specialists that cover both theoretical knowledge and practical application of the standards.
- Online learning platforms: Develop comprehensive e-learning modules that allow professionals to learn and stay updated on TEVV standards at their own pace.
- Industry workshops: Conduct regular hands-on workshops where practitioners can learn to apply TEVV standards to real-world scenarios.

6. Tools and Automation:
Encouraging the development of standardized tools and automated processes to support TEVV activities can significantly enhance efficiency and consistency.

- Open-source initiatives: Foster the development of open-source tools for implementing various aspects of TEVV, encouraging community contributions and widespread adoption.
- Automated testing frameworks: Develop standardized automated testing frameworks that can be easily integrated into digital twin development processes.
- Reporting tools: Create tools that automate the generation of standardized TEVV reports to ensure consistency and reduce the administrative burden on practitioners.

Standardized TEVV approaches provide several key benefits:
Improved Reliability: By following consistent, well-defined processes, the reliability of digital twins can be significantly enhanced, leading to more accurate predictions and simulations.

Enhanced Interoperability: Standardized approaches make it easier for digital twins from different vendors or industries to interact and share data, promoting broader integration and utility.

Accelerated Development: With clear guidelines and standardized processes, the development cycle for digital twins can be streamlined, reducing time-to-market and development costs.

Increased Trust: Rigorous, standardized TEVV processes build confidence in digital twin technology among stakeholders, from end-users to regulators.

Facilitated Compliance: Alignment with regulatory frameworks makes it easier for organizations to demonstrate compliance with industry-specific regulations.

Continuous Improvement: The framework for ongoing review and update of standards ensures that TEVV processes evolve alongside advancements in digital twin technology and changing industry needs.

As digital twins become increasingly prevalent across various sectors, from manufacturing and healthcare to urban planning and aerospace, the importance of standardized TEVV approaches cannot be overstated. These standards will play a crucial role in unlocking the full potential of digital twin technology, driving innovation, and ensuring the reliability and effectiveness of these powerful virtual representations.

**Operationalizing the digital twin TEVV**

1. Planning Phase
   - Define TEVV objectives and scope
   - Identify key stakeholders
   - Establish evaluation criteria and metrics
   - Develop TEVV plan and timeline

2. Design Phase
   - Create test cases and scenarios
   - Design evaluation protocols
   - Develop verification checklists
   - Establish validation criteria

3. Execution Phase
   - Conduct testing activities
   - Perform evaluations
   - Execute verification procedures
   - Carry out validation assessments

4. Analysis Phase
   - Analyze test results
   - Evaluate performance metrics
   - Verify compliance with requirements
   - Validate against real-world data

5. Reporting Phase

- Document findings and observations
- Generate TEVV reports
- Provide recommendations for improvement
- Communicate results to stakeholders

6. Continuous Improvement Phase
   - Implement corrective actions
   - Update digital twin based on TEVV results
   - Refine TEVV processes
   - Conduct periodic reassessments

**Standardized Metrics and KPIs**
To enable consistent evaluation across different digital twins, a set of standardized metrics and Key Performance Indicators (KPIs) should be established. Table 4 summarizes the principal metric categories and specific quantitative measures used to assess digital twin systems within the TEVV framework.

***Table 4: Standardized Evaluation Metrics for Digital Twin TEVV***

| Metric Category | Specific Metrics |
|---|---|
| Accuracy Metrics | • Mean Absolute Error (MAE)<br>• Root Mean Square Error (RMSE)<br>• Coefficient of Determination ($R^2$)<br>• Normalized Root Mean Square Error (NRMSE)<br>• F1 Score |
| Performance Metrics | • Response Time<br>• Throughput<br>• Resource Utilization<br>• Scalability Factor |
| Reliability Metrics | • Mean Time Between Failures (MTBF)<br>• Availability Percentage<br>• Error Rate<br>• Fault Tolerance Index |
| Usability Metrics | • System Usability Scale (SUS) Score<br>• Task Completion Rate<br>• Time on Task<br>• User Satisfaction Rating |
| Value Metrics | • Return on Investment (ROI)<br>• Cost Savings Percentage<br>• Operational Efficiency Improvement<br>• Decision Quality Index |

**Standardized Reporting Templates**

Developing standardized reporting templates ensures consistency in documenting TEVV results across different digital twin projects:

1. Executive Summary
2. Project Overview
3. TEVV Objectives and Scope
4. Methodology
5. Test Results and Observations
6. Evaluation Findings
7. Verification Summary
8. Validation Outcomes
9. Performance Metrics
10. Identified Issues and Risks
11. Recommendations
12. Conclusion
13. Appendices (Test Cases, Raw Data, etc.)

**Certification Process**
Establishing a certification process for digital twins can help ensure adherence to standardized TEVV practices:

1. Pre-assessment: Review of digital twin documentation and TEVV plan
2. TEVV Execution: Conducted by certified assessors or third-party organizations
3. Compliance Evaluation: Assessment against standardized criteria and benchmarks
4. Reporting: Detailed report of findings and compliance status
5. Certification Decision: Awarding of certification based on meeting required standards
6. Periodic Reassessment: Regular reviews to maintain certification status

**Ethical Considerations in Digital Twin TEVV**
As digital twins become more prevalent and influential in decision-making processes, it is crucial to address the ethical implications of their development, deployment, and use (Sel et al., 2025). Below are key ethical considerations that should be integrated into the TEVV process for digital twins.

Privacy and Data Protection
Digital twins often rely on vast amounts of data, including potentially sensitive information about individuals or proprietary business processes. TEVV processes must ensure:

- Data anonymization and de-identification techniques are properly implemented and tested
- Compliance with data protection regulations (e.g., GDPR, CCPA) is verified
- Data access controls and encryption methods are thoroughly evaluated
- Privacy impact assessments are conducted as part of the validation process

Fairness and Non-discrimination

Digital twins used in decision-making processes must be evaluated for potential harmful biases:

- Test for disparate impact across different demographic groups
- Verify that training data is representative and free from historical biases
- Evaluate the fairness of outcomes in various scenarios
- Implement and assess harmful bias mitigation techniques

Transparency and Explainability

The complexity of digital twins can make their decision-making processes opaque. TEVV should address:

- Verification of model interpretability methods
- Evaluation of explanation quality for key stakeholders
- Testing of audit trail and logging mechanisms
- Validation of transparency in data sources and model assumptions

Accountability and Liability

As digital twins influence real-world decisions, clear lines of accountability must be established:

- Verify mechanisms for tracking decisions made based on digital twin outputs
- Evaluate the allocation of responsibility between human operators and automated systems
- Test scenarios involving potential failures or errors
- Validate processes for human oversight and intervention

Environmental Impact

The computational resources required for complex digital twins can have significant environmental implications:

- Assess energy efficiency and carbon footprint of digital twin operations
- Evaluate strategies for optimizing resource usage
- Verify that environmental impact is considered in decision-making processes
- Validate potential environmental benefits against computational costs

Ethical Use and Misuse Prevention

TEVV processes should consider the potential for misuse or unintended consequences:

- Evaluate safeguards against malicious use of digital twins
- Test for vulnerabilities that could lead to system manipulation
- Verify ethical guidelines are integrated into the digital twin's operation
- Validate the alignment of digital twin objectives with broader societal values

Informed Consent and Stakeholder Engagement

When digital twins represent or impact individuals or communities:

- Verify processes for obtaining informed consent
- Evaluate mechanisms for stakeholder feedback and participation
- Test communication methods for explaining digital twin implications
- Validate the incorporation of diverse perspectives in digital twin development

## Case Studies

To illustrate the application of the proposed TEVV framework, three example case studies from different domains are presented:

### Case Study 1: Manufacturing Digital Twin

A large automotive manufacturer implements a digital twin of their production line to optimize efficiency and predict maintenance needs.

TEVV Approach:

1. Testing:
    - Unit testing of individual machine models
    - Integration testing of production line components
    - System testing under various production scenarios
    - Simulation testing with historical data

2. Evaluation:
    - Performance evaluation of prediction accuracy
    - Usability evaluation with production managers
    - Value assessment through cost savings and downtime reduction

3. Verification:
    - Requirements verification against production specifications
    - Data verification of sensor inputs and historical records
    - Model verification of physical process representations

4. Validation:
    - Empirical validation against actual production metrics
    - Predictive validation of maintenance forecasts
    - Operational validation through a pilot implementation

Outcomes:

- Identified and corrected discrepancies in machine interaction models
- Improved prediction accuracy for maintenance needs by 15%
- Validated cost savings of 8% through optimized production scheduling
- Uncovered potential privacy issues in employee productivity tracking

### Case Study 2: Healthcare Digital Twin

A hospital develops a digital twin of patient flow to optimize resource allocation and improve emergency response times.

TEVV Approach:

1. Testing:

- Unit testing of patient intake models
- Integration testing of department interactions
- System testing of hospital-wide patient flow
- Simulation testing with various emergency scenarios

2. Evaluation:
   - Performance evaluation of wait time predictions
   - Usability evaluation with hospital staff
   - Value assessment through improved patient outcomes

3. Verification:
   - Requirements verification against hospital protocols
   - Data verification of patient records and resource availability
   - Model verification of triage and treatment processes

4. Validation:
   - Empirical validation against actual patient flow data
   - Predictive validation of resource needs during peak times
   - Operational validation through staged emergency drills

Outcomes:
- Identified bottlenecks in patient transfer between departments
- Improved emergency response time prediction accuracy by 22%
- Validated 10% reduction in average patient wait times
- Uncovered potential fairness issues in resource allocation algorithms

Case Study 3: Financial/Transactional Digital Twin
A global bank implements a digital twin of its trading and risk management systems to optimize operations, enhance fraud detection, and improve regulatory compliance.

TEVV Approach:
1. Testing:
   - Unit testing of individual financial product models
   - Integration testing of trading platform components
   - System testing under various market scenarios
   - Simulation testing with historical market data and stress test scenarios

2. Evaluation:
   - Performance evaluation of transaction processing speed and accuracy
   - Usability evaluation with traders and risk managers
   - Value assessment through improved risk management and regulatory compliance

3. Verification:
   - Requirements verification against financial regulations and internal policies

- Data verification of market feeds, transaction records, and customer data
- Model verification of risk assessment and pricing algorithms

4. Validation:
    - Empirical validation against actual trading outcomes and risk exposures
    - Predictive validation of market risk forecasts and fraud detection
    - Operational validation through parallel runs with existing systems

Outcomes:
- Identified and corrected discrepancies in complex derivative pricing models
- Improved real-time risk assessment accuracy by 18%
- Validated 25% reduction in false positive fraud alerts
- Uncovered potential ethical issues in algorithmic trading strategies
- Demonstrated 30% improvement in regulatory reporting efficiency

Ethical Considerations:
- Ensured robust data anonymization for customer transaction data
- Implemented fairness checks in credit risk models to prevent discrimination
- Enhanced transparency of AI-driven trading decisions for regulatory scrutiny
- Established clear accountability protocols for automated vs. human-driven transactions
- Evaluated the environmental impact of high-frequency trading simulations

Case Study 4: Urban Planning Digital Twin
A city develops a digital twin to model traffic flow, air quality, and energy usage for urban planning decisions.

TEVV Approach:

1. Testing:
    - Unit testing of traffic signal models
    - Integration testing of transportation network components
    - System testing of city-wide interactions
    - Simulation testing with various urban development scenarios

2. Evaluation:
    - Performance evaluation of traffic prediction accuracy
    - Usability evaluation with city planners and policymakers
    - Value assessment through improved quality of life metrics

3. Verification:
    - Requirements verification against urban planning guidelines
    - Data verification of sensor networks and historical city data
    - Model verification of environmental impact calculations

4. Validation:

- Empirical validation against real-time city metrics
- Predictive validation of long-term urban development outcomes
- Operational validation through pilot projects

**Limitations**

The proposed TEVV framework for digital twins has several limitations that should be acknowledged:

Complexity and Resource Intensity

- Implementing the full TEVV framework may be resource-intensive, particularly for smaller organizations or less complex digital twins.
- The depth of testing and validation required may extend development timelines and increase costs.

Evolving Technology

- As digital twin technology rapidly evolves, the TEVV framework may need frequent updates to remain relevant.
- New types of digital twins may emerge that require novel testing approaches not covered by this framework.

Domain-Specific Challenges

- While the framework aims to be broadly applicable, certain industries may face unique challenges that require specialized TEVV methods.
- Highly regulated industries may need additional validation steps beyond those outlined in the general framework.

Data Limitations

- The effectiveness of TEVV processes is heavily dependent on the quality and quantity of available data.
- In some cases, obtaining sufficient real-world data for comprehensive validation may be challenging or impossible.

Human Factors

- The framework may not fully capture the complexities of human interaction with digital twins, particularly in decision-making processes.
- Usability evaluations may be subjective and vary across different user groups.

Interoperability Issues

- As digital twins become more interconnected, testing their interactions with other systems and twins may become increasingly complex.
- Standardization efforts across different industries and vendors may lag behind technological advancements.

**Broader Impacts**

The implementation of standardized TEVV approaches for digital twins has far-reaching implications across various sectors and society as a whole:

Economic Impacts
- Improved reliability and performance of digital twins could lead to significant cost savings and efficiency gains across industries.
- The growth of the digital twin market may create new job opportunities in TEVV specialization.
- Smaller businesses may face challenges in adopting comprehensive TEVV practices, potentially widening the technology gap.

Technological Advancement
- Standardized TEVV practices could accelerate the development and deployment of more sophisticated digital twins.
- Increased confidence in digital twin technology may spur further innovation and investment in related fields like AI and IoT.

Environmental Considerations
- Digital twins could play a crucial role in optimizing resource usage and reducing waste across various industries (Dai et al., 2024; Luo et al., 2022).
- However, the energy consumption of complex digital twins and their supporting infrastructure needs careful consideration.
- The development of Earth-scale digital twins is poised to transform environmental monitoring, climate modeling, and disaster response by providing integrated, real-time virtual representations of natural systems (Hazeleger et al., 2024; Blair, 2025).

Social and Ethical Implications
- Improved digital twins could enhance public services, urban planning, and healthcare outcomes.
- Ethical use of digital twins in decision-making processes will require ongoing scrutiny and governance.
- Privacy concerns related to data collection and use in digital twins will need continuous attention.

Education and Workforce Development
- The need for specialized skills in digital twin TEVV will likely influence educational curricula and professional training programs.
- Interdisciplinary collaboration may increase as digital twins bridge various fields of expertise.

Policy and Regulation
- Standardized TEVV practices may inform new regulations and standards for digital twin development and deployment.
- International cooperation may be necessary to establish global standards for digital twin TEVV.

Research Directions

- The framework provides a foundation for further research into TEVV methodologies specific to digital twins.
- It may stimulate new areas of study in fields like systems engineering, data science, and human-computer interaction.

**Conclusion**

The proposed framework for testing, evaluation, verification, and validation (TEVV) of digital twins represents a step towards ensuring the reliability, accuracy, and ethical implementation of this transformative technology. Establishing standardized approaches to TEVV can cultivate trust in digital twins across various industries and applications.

The framework addresses the unique challenges posed by digital twins, including their dynamic nature, complex interactions with physical systems, and potential for far-reaching impacts on decision-making processes. It provides a structured methodology that encompasses rigorous testing procedures, multi-faceted evaluation techniques, thorough verification processes, and robust validation methods.

Key strengths of the framework include its adaptability to different types of digital twins, from component-level models to enterprise-wide systems. The emphasis on continuous improvement and ethical considerations ensures that the framework can evolve alongside technological advancements and societal needs.

However, the limitations identified, such as resource intensity and the challenges of keeping pace with rapidly evolving technology, highlight the need for ongoing refinement and adaptation of TEVV practices. Future work should focus on addressing these limitations and expanding the framework to encompass emerging types of digital twins and novel applications.

The broader impacts of implementing standardized TEVV approaches for digital twins are profound and multifaceted. From economic benefits and technological advancements to social implications and environmental considerations, the ripple effects of improved digital twin reliability and performance will be felt across society.

As we move forward, it will be important to maintain a balance between rigorous TEVV practices and the need for innovation and flexibility in digital twin development. Collaboration between industry, academia, and regulatory bodies will be essential in refining and implementing these standards.

Ultimately, this TEVV framework provides a foundation for developing more trustworthy, effective, and ethically sound digital twins. Adherence to these standardized practices can help unlock the full potential of digital twin technology to drive innovation, improve decision-making, and address complex challenges across various domains. As digital twins continue to shape our world, confirming their accuracy, reliability, and responsible use through comprehensive TEVV processes is not just a technical necessity but a societal imperative (David & Bork, 2023).

## References

1. Fuller, A., Fan, Z., Day, C., Barlow, C. *Digital Twin: enabling technologies, challenges and open research*. (2020). IEEE Journals & Magazine | IEEE Xplore. https://ieeexplore.ieee.org/document/9103025

2. Tao, F., Zhang, H., Liu, A., Nee, A. *Digital twin in industry: State-of-the-Art*. (2019, April 1). IEEE Journals & Magazine | IEEE Xplore. https://ieeexplore.ieee.org/document/8477101

3. Jones, D., Snider, C., Nassehi, A., Yon, J., & Hicks, B. (2020). Characterising the Digital Twin: A systematic literature review. *CIRP Journal of Manufacturing Science and Technology*, *29*, 36–52. https://doi.org/10.1016/j.cirpj.2020.02.002

4. Barricelli, B., Casiraghi, E., Fogli, D. (2019). *A survey on Digital Twin: Definitions, characteristics, applications, and design implications*. (2019). IEEE Journals & Magazine | IEEE Xplore. https://ieeexplore.ieee.org/document/8901113

5. Sel, K., Hawkins-Daarud, A., Chaudhuri, A., Osman, D., Bahai, A., Paydarfar, D., Willcox, K., Chung, C., & Jafari, R. (2025). Survey and perspective on verification, validation, and uncertainty quantification of digital twins for precision medicine. *Npj Digital Medicine*, *8*(1). https://doi.org/10.1038/s41746-025-01447-y

6. Hua, E., Lazarova-Molnar, S., Francis, D. *Validation of digital twins: challenges and opportunities*. (2022). IEEE Conference Publication | IEEE Xplore. https://ieeexplore.ieee.org/document/10015420

7. Thelen, A., Zhang, X., Fink, O., Lu, Y., Ghosh, S., Youn, B. D., Todd, M. D., Mahadevan, S., Hu, C., & Hu, Z. (2022). *A comprehensive review of Digital Twin -- Part 2: Roles of uncertainty quantification and Optimization, a battery Digital Twin, and Perspectives*. *arXiv* (Preprint No. 2208.12904). https://arxiv.org/abs/2208.12904

8. Lugaresi, G., Gangemi, S., Gazzoni, G., Matta, A. (2023). Online Validation of Digital Twins for Manufacturing Systems. *Computers in Industry*, 150. https://doi.org/10.1016/j.compind.2023.103942

9. Somers, R. J., Douthwaite, J. A., Wagg, D. J., Walkinshaw, N., & Hierons, R. M. (2022). Digital-twin-based testing for cyber–physical systems: A systematic literature review. *Information and Software Technology*, *156*, 107145. https://doi.org/10.1016/j.infsof.2022.107145

10. Huang, L., Varshney, L. R., & Willcox, K. E. (2024). *Formal Verification of Digital Twins with TLA and Information Leakage Control*. arXiv.org. https://arxiv.org/abs/2411.18798

11. Lu, H., Zhang, L., Wang, K., Huang, Z., Cheng, H., & Cui, J. (2024). A Framework for the Credibility Evaluation of Digital Twins. In *Simulation foundations, methods and applications* (pp. 69–93). https://doi.org/10.1007/978-3-031-69107-2_4

12. Gao, T., Chen, L., Zhang, X., Guo, J., & Ni, D. (2025). Credibility Assessment for Digital Twins in Vehicle-in-the-Loop Test Based on Information Entropy. *Sensors*, *25*(5), 1372. https://doi.org/10.3390/s25051372

13. Shao, G., Kibira, D., Frechette, S. (2024). *Credibility consideration for digital twins in manufacturing*. (NIST SP 1500-21). National Institute of Standards and Technology. https://tsapps.nist.gov/publication/get_pdf.cfm?pub_id=957417

14. Katsoulakis, E., Wang, Q., Wu, H., Shahriyari, L., Fletcher, R., Liu, J., Achenie, L., Liu, H., Jackson, P., Xiao, Y., Syeda-Mahmood, T., Tuli, R., & Deng, J. (2024). Digital twins for health: a scoping review. *Npj Digital Medicine*, *7*(1). https://doi.org/10.1038/s41746-024-01073-0

15. Venkatesh, K. P., Raza, M. M., & Kvedar, J. C. (2022). Health digital twins as tools for precision medicine: Considerations for computation, implementation, and regulation. *Npj Digital Medicine*, *5*(1). https://doi.org/10.1038/s41746-022-00694-7

16. Bian, J., Huang, Y., Dai, H., Xu, J., Wei, R., Sun, L., Guo, Y., & Guo, J. (2024). Evolution of digital twins in precision health applications: a scoping review study. *Research Square (Research Square)*. https://doi.org/10.21203/rs.3.rs-4612942/v1

17. Shen, Md., Chen, Sb. & Ding, Xd. The effectiveness of digital twins in promoting precision health across the entire population: a systematic review. *npj Digit. Med.* 7, 145 (2024). https://doi.org/10.1038/s41746-024-01146-0

18. Sun, T., He, X., Song, X., Shu, L., & Li, Z. (2022). The Digital Twin in Medicine: A Key to the Future of Healthcare?. *Frontiers in medicine*, *9*, 907066. https://doi.org/10.3389/fmed.2022.907066

19. Karabulut, E., Pileggi, S. F., Groth, P., & Degeler, V. (2023). Ontologies in digital twins: A systematic literature review. *Future Generation Computer Systems*, *153*, 442–456. https://doi.org/10.1016/j.future.2023.12.013

20. Wilson, F., Hurley, R., Maxwell, D., McLellan, J., & Beverley, J. (2024). *Foundations for digital twins*. arXiv.org. https://arxiv.org/abs/2405.00960

21. Petrova-Antonova, D., & Ilieva, S. (2020). Digital twin modeling of smart cities. In *Advances in intelligent systems and computing* (pp. 384–390). https://doi.org/10.1007/978-3-030-55307-4_58

22. Qi, Q., Tao, F., Hu, T., Anwer, N.n Liu, A., Wei, Y., Wang, L., Nee, A. Enabling technologies and tools for digital twin. (2021). *Journal of Manufacturing Systems* 58(Part B). 3-21. https://doi.org/10.1016/j.jmsy.2019.10.001

23. McClellan, A., Lorenzetti, J., Pavone, M., & Farhat, C. (2022). A physics-based digital twin for model predictive control of autonomous unmanned aerial vehicle landing. *Philosophical Transactions of the Royal Society a Mathematical Physical and Engineering Sciences*, *380*(2229). https://doi.org/10.1098/rsta.2021.0204

24. Cimino, C., Negri, E., & Fumagalli, L. (2019). Review of digital twin applications in manufacturing. *Computers in Industry*, *113*, 103130. https://doi.org/10.1016/j.compind.2019.103130

25. Kapteyn, M., Knezevic, D., Huynh, D., Tran, M., & Willcox, K. (2020). Data-driven physics-based digital twins via a library of component-based reduced-order models. *International Journal for Numerical Methods in Engineering*, *123*(13), 2986–3003. https://doi.org/10.1002/nme.6423

26. Ding, X., & Xing, J. (2025). Modeling and implementation of a real-time digital twin for the Stewart platform with real-time trajectory computation. *PeerJ Computer Science*, *11*, e2892. https://doi.org/10.7717/peerj-cs.2892

27. Deantoni, J., Muñoz, P., Gomes, C., Verbrugge, C., Mittal, R., Heinrich, R., Bellis, S. & Vallecillo, A. (2025). Quantifying and combining uncertainty for improving the behavior of Digital Twin Systems. *at - Automatisierungstechnik*, *73*(2), 81-99. https://doi.org/10.1515/auto-2024-0036

28. Wagg, D. J., Worden, K., Barthorpe, R. J., & Gardner, P. (2020). Digital Twins: State-of-the-Art and future Directions for modeling and simulation in Engineering Dynamics Applications. *ASCE-ASME Journal of Risk and Uncertainty in Engineering Systems Part B Mechanical Engineering*, *6*(3). https://doi.org/10.1115/1.4046739

29. David, I., Bork, D. *Towards a taxonomy of digital twin evolution for technical sustainability*. (2023). IEEE Conference Publication | IEEE Xplore. https://ieeexplore.ieee.org/document/10350387

30. Blair, G. (2025). Digital twins of the natural environment. *Patterns 2(10).* https://doi.org/10.1016/j.patter.2021.100359

31. Hazeleger, W., Aerts, J. P. M., Bauer, P., Bierkens, M. F. P., Camps-Valls, G., Dekker, M. M., Doblas-Reyes, F. J., Eyring, V., Finkenauer, C., Grundner, A., Hachinger, S., Hall, D. M., Hartmann, T., Iglesias-Suarez, F., Janssens, M., Jones, E. R., Kölling, T., Lees, M., Lhermitte, S., van Nieuwpoort, R., Pahker, A., Pellicer-Valero, O., Pijpers. F., & Vossepoel, F. C. (2024). Digital twins of the Earth with and for humans. *Communications Earth & Environment*, *5*(1). https://doi.org/10.1038/s43247-024-01626-x

32. Tang, Z., Zhuang, D., & Zhang, J. (2025). Evaluation framework for domain-specific digital twin platforms. *Scientific Reports*, *15*(1). https://doi.org/10.1038/s41598-024-82154-8

33. Cardoso, F. S., De Souza Araújo, T., Rohrich, R. F., & De Oliveira, A. S. (2025). Modeling, evaluation and metrics performance of the SyncLMKD in distributed kinematics variations. *Scientific Reports*, *15*(1). https://doi.org/10.1038/s41598-024-84997-7

34. Shao, G. (2024). Manufacturing Digital Twin Standards. *Association for Computing Machinery*, MODELS Companion '24: Proceedings of the ACM/IEEE 27th International Conference on Model Driven Engineering Languages and Systems. 370–377. https://doi.org/10.1145/3652620.3688250

35. Drobnjakovic, M., Shao, G., Nikolov, A., Kulvatunyou, B., Frechette, S., Srinivasan., V. National Institute of Standards and Technology. (2023). *Towards ontologizing a digital twin framework for manufacturing*. (NIST Internal Report 8356). https://tsapps.nist.gov/publication/get_pdf.cfm?pub_id=936637

36. Liu, M., Fang, S., Dong, H., & Xu, C. Review of digital twin about concepts, technologies, and industrial applications. (2021). *Journal of Manufacturing Systems 58(Part B).* https://doi.org/10.1016/j.jmsy.2020.06.017

37. Pronost, G., Mayer, F., Marche, B., Camargo, & M., Dupont, L. *Towards a Framework for the Classification of Digital Twins and their Applications*. (2021). IEEE Conference Publication | IEEE Xplore. https://ieeexplore.ieee.org/abstract/document/9570114

38. Flammini, F. (2021). Digital twins as run-time predictive models for the resilience of cyber-physical systems: a conceptual framework. *Philosophical Transactions of the Royal Society a Mathematical Physical and Engineering Sciences*, *379*(2207), 20200369. https://doi.org/10.1098/rsta.2020.0369

39. Dai, C., Cheng, K., Liang, B., Zhang, X., Liu, Q., & Kuang, Z. (2024). Digital twin modeling method based on IFC standards for building construction processes. *Frontiers in Energy Research*, *12*. https://doi.org/10.3389/fenrg.2024.1334192

40. Luo, F., Feng, S., Yang, Y., Ao, Z., Li, X., & Chai, Y. (2022). Ontology modeling method applied in simulation modeling of distribution network time series operation. *Frontiers in Energy Research*, *10*. https://doi.org/10.3389/fenrg.2022.935026

41. Bitencourt, J., Wooley, A., & Harris, G. (2024). Verification and validation of digital twins: a systematic literature review for manufacturing applications. *International*

*Journal of Production Research*, 1–29. https://doi.org/10.1080/00207543.2024.2357741

42. Ammar, M., Mousavi, A., & Al-Raweshidy, H. (2024). Physics Based Digital Twin Modelling from Theory to Concept Implementation Using Coiled Springs Used in Suspension Systems. *Tech Science Press*. https://doi.org/10.32604/dedt.2023.044930